\documentclass[preprint,showpacs,preprintnumbers,amsmath,amssymb]{revtex4}

\usepackage{graphicx}
\usepackage{dcolumn}
\usepackage{bm}

\begin{document}

\preprint{YITP-05-67}
\preprint{OIQP-05-14}

\title{
Non-existence of Irreversible Processes 
in Compact Space Time}

\author{Holger B. Nielsen}
\affiliation{
Niels Bohr Institute, \\ 
University of Copenhagen \\ 
17 Blegdamsvej, Copenhagen $\phi$, Denmark}

\author{Masao Ninomiya }%
 \altaffiliation[Also at ]{Okayama Institute for Quantum
Physics, Kyoyama-cho 1-9, Okayama City 700-0015, Japan.}
\affiliation{%
Yukawa Institute for Theoretical Physics, \\
Kyoto University \\ 
Kyoto 606-8502, Japan
}%


\begin{abstract}
It is shown that if physical space time were truly compact there would only be of the
order of one solutions to the classical field equations with a weighting
to be explained. 
But that would not allow any peculiar choice of initial
conditions that could support a non-trivial second law of
thermodynamics.
We present a no-go theorem: 
Irreversible processes would be extremely
unlikely to occur for the almost unique solution for the
intrinsically compact space time world, 
although irreversible processes are well known to occur in general.
What we here assume -- compact space time -- excludes that universe
could exist eternally. 
In other word if universe stays on forever (i.e. non-compact in time direction) 
our no-go theorem is not applicable.
\end{abstract}

\maketitle

\section{Introduction}
It has been suggested by Hartle and Hawking that the initial state of
the universe -- the wave function of the universe -- be given by the
``no boundary postulate''.
\cite{1}
The idea is that the wave function be given by a functional integral
over the field configurations on a space time manifold.
If the model is used only to give a wave function there is 
made an assumption about a kind of past, but no
assumption about the future that also ends without boundary.
It is not in such a wave function for the universe formulation assumed
should be valid for the future
development.
It would, however, seem natural from the spirit of the no boundary
assumption that it should be valid for both future and past.
We have also recently discussed a model with in a slightly
special way of compactified space time -- or rather just time --
\cite{3, 4, 5, 6, 7}, in which the time axis  be an $S^1$-circle.
In these articles we discussed the possibility of the time axis
with $S^1$-topology, which is equivalent to consider a world with
an intrinsic periodicity.
The major problem for such a model which we found was that entropy
would be constant as a function of time, i.e. there would be no
irreversible processes in such a universe. 
Our main concern in the present article is an extension of this result: 

Provided the space time is \underline{compact} there cannot be
irreversible processes, in the classical approximation.

The main point of the present article is indeed that for a compact
space time the number of solutions to the classical equations of
motion becomes essentially of order one 
after an appropriate discretization, or weighted in a
way to be discussed below. 
Then there is no place for making more assumptions
about the special solution to the equations of motion selected by
Nature.
Thus there is no more any possibility for specifying the
initial state which leads for us to present a \underline{no-go theorem}
that will be discussed in detail in section 5 in the present article. 
But such an initial state specification is highly needed to implement
the second law of thermodynamics in a non-trivial way. 
By a non-trivial implementation we here mean one involving
irreversible processes.

In the following section 2, we shall seek as to
how we think about the very general classical field
theories discussed here in a concrete manner, 
and how we cut them off, both by a lattice in
space time and by a lattice in the value space for the fields
(components).
We allude to our philosophy that the equations are
sufficiently complicated such that we can use statistical arguments in
discussing them.
In section 3 we then discuss a few aspects of estimate of the number of 
solutions for a compact space time, but we seek also to put the
effect of non-compact corners of the space time manifold in
perspective.
In section 4 we introduce the concept of macro state developments so as
to be able to discuss entropy and what an irreversible process
involves. 

Then in section 5 we complete the argument that irreversible processes
are extremely unlikely to occur for the ``essentially'' unique
solution to the equations of motion.
In section 6 we conclude and put forward some outlook as to how to get
irreversible processes at all compatible with a sensible micro
physics cosmology.

\section{General classical field theory and cutting it off}

It is the spirit of the present article to perform an extremely
abstract and general discussion about a general classical field theory
with fields defined as functions on a space time manifold as in
general relativity.
The main interest in this article is to make a no go theorem for the
compact space time manifold, but at first we can set
up our formalism and thinking even for manifolds that are not
compact.

To avoid a lot of technical problems and details in connection with
Lyapunov exponents 
\footnote{See e.g. ref.\cite{7}}
which anyway turn out to cancel, we shall work with
a regularized or cut off theory.
Since we work only classically, the usual type of cut off on 
quantum field theory will, however, not be needed. 
Nevertheless we shall consider two types of cut off:
\begin{enumerate}
 \item[1)] We shall think of a latticification of space time by letting the
   classical fields be defined only on a set of sites.
Since the manifold 
has a priori even a complicated topology -- at least we should be able
to have a compact manifold -- we cannot have a completely regular
lattice on the manifold, but that is presumably not needed either for
the very abstract discussions in the present article.

\item[2)] We shall also discretize the set of values that the assumed
   multicomponent field can take on.
\end{enumerate}

After making these two cut off procedures our general classical field
theory has been converted into a set of ``lattice'' point on the
manifold on which multicomponent fields are defined that take values
in a discrete value space.

The equations of motion will be approximated by relations between the
field values on a bunch of a few neighboring sites.

If we imagined that at the continuum field level we introduced
enough conjugate momentum fields so as to make the field equations
become first order partial differential equations, a naive
discretization would lead to the form that we could write the field equations
as a relation between the field components on a bunch of $d+1$
neighboring sites.  Here $d$ is the dimension of the space time
manifold so that in the physically relevant case this dimension would
be $d=4$. Thus we should have relations for bunches of 5
neighboring points.  
Since we are mainly interested in counting
equations and degrees of freedom it is very important that we shall
have just one such relation between a bunch of $d+1$ lattice points
for each lattice point.  That is to say we shall have in the bulk of
the manifold just equally many relations as field components on the
sites.  
More precisely, if we call the number of field components for
the field $ \phi $, say $n$, then there are $n$ field components per
site.
Since there should be also $n$ equations -- i.e. an
$n$-component relation there should be just equally
many relations as variables in the bulk.
This is what is naively taken a discretized
solution set.

In cut off language the statement of the discretized solutions 
looses its interest because any solution is anyway discretized.  
We have however still some meaning attached to the
concept analogous to the fact that there are equally many equations as
variables:  We have a disposition set $F$ for simplicity, with a
finite number of values $l$; cardinal number of $F$ 
$=ktF$ for the multicomponent field at a site.  
If each of the $n$
components could take $k$ values, $l=k^n$.  
Then a $d+1$ site involving relation $R$ is a subset of 
$\underbrace{F \times \cdots \times F}_{d+1} = F^{d+1}$.  
If this relation corresponds to an equation of the deterministic type
in accordance to the equations of motion, we should require that for
given values $y_i$ ($i \not = j$) of all but one in an ordered set
$(y_1, y_2, \cdots, y_{j-1}, x, y_{j+1}, \cdots, y_{d+1}) \in R$ the
component $x$ sitting on the $j$-th place is uniquely
determined from the requirement of the full ordered set of $d+1$
components belonging to the relation $R$.  We both require that such
an $x$ exists and that there be just one $x$ for any 
choice of the $(d+1)-1=d$ other components.  In this case the cardinal
number for the relation $R$ is just 
${{(ktF)^{d+1}} \over {ktF}} = (ktF)^d = l^d$. 
For the just described type of relation between the $d+1$ sites we say that
it reflects one equation.  

It is the view point of the present article that the equations of
motion are so complicated and have so many parameters -- coupling
constants etc. -- that we effectively consider them random.  There is
though the exception that we shall below 
assume that the random elements in the equations of motion do not mix 
the different macro states, when we study macro states
and macro restrictions in order to be able to discuss the concept of
irreversible processes.  
Return we shall to this exception in the randomness below in section 4.

Inside the macro restrictions imposed we can consider the
relations as random.  That is to say that we in principle let $R$ be
chosen as a random one among all those obeying the properties
described from the macro description requirements. 

Such considerations are, however, only used in the mild form 
in such the case as we use them as an excuse to simply make statistical
considerations  to estimate e.g. numbers of solutions.  But it
should be kept in mind that such estimations will work in almost all
cases and that it is thus a very mild assumption to the
validity of such arguments.

\section{Number of solutions}

In the case in which there are just equally many equations as
variables as is the case for 
one field, i.e. the equations with the same number of components as the
fields per lattice point, we have already seen that the solutions
are discrete even when we do not discretize the field value space.  

When we instead think of the discretized field value space 
we can use this coincidence of number of
equations and variables to conclude that the number of
solutions will at most be of order unity, if there are solutions at
all.  The argument is as follows:

If we have the number of lattice points $\sharp L$ on which the field $\phi$ is
defined and takes values in a disposition space 
$F$ with $l = \sharp F = ktF$ elements, 
there exist a priori $l^{\sharp L}$ different field functions. 
Here $kt$ denotes cardinal number of $F$. 
We impose each time the field equation relation
involving as above described $d+1$ lattice points and only keep
those field configurations satisfying the imposed relation, we reduce
the number of field configurations by a factor $l$.  The relation $R$
only allow one out of $l$ configurations on the $d+1$ sites 
involved.  After imposing $\sharp L$ equations, which means one per site we
thus get reduced the original number of field functions on the lattice
$l^{\sharp L}$ by a factor $l^{\sharp L}$ leaving a number of
solutions of the order of 
${{l^{\sharp L}} \over {l^{\sharp L}}} \simeq 1$. 
This means that it could easily be zero, i.e. no solutions at
all, or it could be few.  It could, however, not likely be a 
huge number unless there is some systematic regularity, which is
really what we thought to exclude by our assumptions in the previous section 2.

This result is of major importance for the present article because it
means that approximately and statistically there is just one solution
to the equations of motion in latticized and in value or disposition
space discretized model by assuming that when there is a finite number
of lattice points as it corresponds to a compact space time.

If the lattice $L$ is allowed to become infinite or if the value space
is allowed to be infinite the above simple counting may no longer be
trustable.  

What could be especially importance would be if in some direction,
typically in the $A \rightarrow \infty $ direction there occurs an
infinite series of lattice points and if even say the universe
expanded so that for cut off in time by only including $A \leq T$
where $T$ is the infrared cut off in time, there become more and more
lattice points on the infrared cut off border.  In this case of the 
$A \leq T$ cut off with even expanding universe there will, 
for larger and larger $T$, be unavoidably more and more sets of
the $d+1$ lattice points associated with equation of motion which will
be cut into pieces by the infrared cut off $A \leq T$.  That is to say
there will be more of these $d+1$ local subsets of the lattice
that have some of their elements with $A \leq T$ and some with $A >
T$.  These ``cut to pieces'' equations of motion elements (we may call
such equations as equation of motion elements) can thus not be imposed on
the infrared cut off theory.  But then it means that they are lacking as
equation of motion elements whereas so to speak the corresponding
sites may well fall inside the included region $A \leq T$.  It is easy
to see that there is good reasons for expecting -- statistically at
least -- that there will be more sites in the included region 
$A \leq T$ than fully included sets of $d+1$ sites associated
with equations of motion elements.  This then means that the
dimensionality of the solution space -- in the continuum case -- will
go up and up with $T$ proportional to the space volume of the
universe at time $T$. 
The latter will be proportional to the
number of cut $d+1$ sets associated with the elements of equations of
motion.

In the discretized or cut off language we will instead find that the
number of solutions for the infrared cut off space time no longer is
of order unity.  Rather it goes as
\begin{eqnarray}
l^{\sharp(L \cap \{t<T\})} - \sharp^{uncut~equations} 
\nonumber\\~~~
\sim l^{constant \cdot ``Space~vol~at~t=T"}
\end{eqnarray}
Such a factor could be more significant than what we can obtain by
uncertainties in the argument for the compact space time.

It should, however, be kept in mind that our level of intention with
respect to the accuracy of the estimates for number of solutions is low.  
The reason is that we intend to compare the logarithms of such
numbers with entropies measured in Boltzmann constant $k$ unit.  
In this natural unit $k$ the entropies are very large numbers
-- we could say Avogadro's number sizes -- and the exponentials of
them become even much more huge.  
If we only care for such an accuracy
level every sensible number is of order unity.  

We see that it is at least not excluded in the above arguments 
that a universe keeping to expand 
or just staying huge into an infinite future that needs an
infrared cut off can cause molester of so many elements of
equation of motion. 

Thus our counting solutions to be essentially of order unity indeed
depends strongly on compactness of the space time.
In fact at least a big universe existing into an infinite future time axis 
$A \rightarrow \infty $ is excluded.

\section{Macro state, and what is the significance of
irreversible process?}
\subsection{Why do we make classical thinking?}

Since the main point of the present article is that the empirical
occurrence of irreversible processes such as friction or heat
conduction is not compatible with a classical field theory with
compact space time in the setting introduced above, we need to
put the concept of irreversible processes into our language.  
Usually
one describes the processes which may possibly be irreversible in terms
of thermodynamics states, which we may call macro states, meaning that
they are states of whole macroscopic systems described by extremely fewer
parameters than the ones needed for the micro degrees of freedom.  That is
to say we here take the point of view that corresponding to each or at
least the most important fundamental states of the micro degrees of
freedom there is a macro state. 
Typically there will be a huge number of
different micro states that correspond to the same macro state.
Usually one thinks of dimensions of the quantum mechanical Hilbert
space, when one talks about the micro states.  
But we could equally
well think about the micro state as a point in the phase
space, and one could make a latticification of the phase space which
simulates the Heisenberg uncertainty principle by taking the discrete 
points to approximately cover a volume as required by this
uncertainty principle.  
There are several -- or rather two --
different ways of thinking about the micro degrees of freedom as
classical:  

\begin{enumerate}
  \item[1)] 
 We could consider the micro world described by fields obeying
    Klein-Gordon equations and Maxwell equations.  We could even
   think of Dirac or Weyl equations, but we would rather like to ignore
   the Fermion fields to avoid further complications in the present
   description of our ideas and make classical approximations
   for these fields.  This is actually the method of making a
   classical description the most easily matching with the discussion in
    the previous sections.
\item[2)]
 An alternative and different way of making a classical theory
   approximation to the micro world is to describe the various
   elementary particles such as elections, protons, etc as classical
   particles in terms of their positions and momenta and with some
   their internal degrees of freedom.  Especially
   describing the motion of molecules by such classical approximation
   may be a good approximation in some cases.  
\end{enumerate}

If one wants to be very precise one clearly needs to work
quantum mechanically, but our main point is so abstract and our
accuracy requirements so low requirement of accuracy that
presumably even the crudeness of working classically may not matter for
the accuracy sake.

However, philosophically it may open for the danger of throwing away
our main point by clinging too much to quantum mechanics.  In fact it
may be tempting to invoke to an anthropic principle by using that we
already know a lot empirically about that quantum state in which 
many world interpretation \cite{8, 9, 10}
have singled out a component of the wave function a l\'{a} Everett.  
This component could well be one making up at the end an extremely small
part of the original Hartle-Hawking's wave function.  
By so doing one may thus soon end up working with a model that in reality only
uses an extremely unlike probability part of the original 
Hartle-Hawking's wave function.  
This is a bit like assuming that
humanity were created by miracle if just the existence of humanity
selects a suppressed probability amplitude.  

In the classical language which we prefer here to use we can more
clearly see that if the number of solutions with the empirically
correct behavior is much under unity the model truly speaking does not
function.  We should not believe that in a theory leaving zero solution
behaves correctly compared to phenomenology.

Thus to not have all the problems about thinking objects or states only
exist after they are observed, we prefer to use classical thinking 
which is a kind of niche for the present series of works.  
Then also the future will exist no matter 
whether we measure it or not and there will
be a definite future calculable by means 
of the equations of motion from the past or in the case of a
compactified space time even from the equations of motion alone!

\subsection{What is an irreversible process}

The irreversible processes are characterized by the entropy truly
increasing with time, with strict inequality i.e. $\dot{S} > 0$.  
Here we
suggested to use a definition of entropy as the logarithm of the
number of micro states.  
Let us say we have put a lattice into the phase space and then the micro state
in the above definition of entropy 
corresponds to the macro state which then the
macro state contains the micro state and is thereby realized.  
Here the irreversible process means that the universe develops
from a macro state with smaller entropy into the one with bigger.  
This development is in conformity with equations of motion.  
Generally classical equations of motion correspond
actually to the unfolding as time goes on by a canonical
transformation and as such conserves the phase space volume.  
That is to say that a low entropy macro state 
develops under the
irreversible process into a macro state with a much bigger phase
space volume.  
Hugely bigger because the phase space volume of the
high entropy macro state is the exponential of this high entropy and
entropies in general measured in the Boltzmann constant as unit are
already very big numbers.  But the phase space volume is
conserved by the time development as is given by the fact that 
the equations of motion
conserve the phase space volume because it were canonical transformation. 
We must understand that in the possible micro states 
often the irreversible process has taken place.
The system can now be in is only a very
tiny subvolume of the whole high entropy macro state phase space
volume.  To the discretization introduced there corresponds to the one
of the phase space volume into points in phase space.
In this cut off language the irreversible process implies then that
there is only a very tiny number of micro states, though still huge
relative to the even more huge total number of micro states in the
high entropy macro state into which there is any chance for the
system to go.  
The restriction to this tiny subset comes in because of the
system having to come through the low entropy macro state.  
This
means that only an exceedingly tiny part of the micro states in the
high entropy macro state are really unstable if they agree with
the experiment or phenomenological observation of the
irreversible process that took place
and thus the solution -- if there is only one as we derived in the
compact space time case ``essentially'' -- has to belong to this tiny
subset.

\section{The crux of our argument: a compact space time universe
is not phenomenologically viable}

\subsection{Random part of Hamiltonian and irreversible process \\
-- a no-go theorem --}
In the view point that there exists 
a random part of the Hamiltonian 
at least as far as the moving around of the micro state inside its macro state 
we shall show in this subsection some conflict between the compact space time
and the existence of irreversible processes.  
This random part of the Hamiltonian will 
be able to push around at random the relatively tiny subset of micro
states inside the high entropy macro state, 
although we assume it restricted not to let the
transitions between macro states be changed by the random part.  
This means that even if
we thought at first that the essentially unique solution -- in case of
compact space time --
belonged to this tiny subset, this property would be spoiled by a
bit of change in the random part of the Hamiltonian.  The main point
is that in the random part of Hamiltonian it becomes
exceedingly unlikely that the ``essentially'' unique solution should
just be one of the tiny subset which could
have come from the low entropy state by equation of motion.  

We may make statement of the crux of the argument in the following:

By varying a bit randomly a part of the Hamiltonian which by itself
are not causing transition between macro states, but they are caused by a past
which is not considered random here, the ``essentially'' unique solution 
is varied naturally in the compact space time case.  
This random variation of the random part most
importantly changes how the micro state is in detail in the era in
which it passes through the macro state, we called the high entropy one.  
But now for the observed irreversible process to occur for the world
developing accordingly to the ``essentially'' unique solution it is
necessary that the latter during the passage of the high entropy
macro state belongs to the tiny subset.  
But that is 
under the influence of the random Hamiltonian shuffling around
exceedingly unlikely.  
This is the crux of our argument:  It is
exceedingly unlikely that the ``essentially'' unique solution will
match to this requirement from the observation of the irreversible
process.  

The conclusion from this then is that a compact space time is not
phenomenologically viable!  Logically of course we made a series of
assumptions and approximations, but it is our point of view that the
extra assumptions were very mild.  By this mildness we mean that the
content in them which were really relevant was correct and just healthy
scientific judgment.

For instance our cut off procedures above are not reliable 
as to whether it really exists in nature -- 
although we would not exclude the possibility that something
close to that could be true -- but we believe that we could with 
some mathematical effort reproduce the essential points by using
either continuum classical physics or even quantum mechanics and quantum
field theory.  Also one should not take literally that we
have a random contribution to the Hamiltonian.  However, to really
invalidate our argument by avoiding this random Hamiltonian philosophy
would mean to have the solution do something that looked
backward in time would be completely miraculous.  
To circumvent the random
Hamiltonian part story we might simply bear in mind that the
``essentially'' unique solution was determined from the equations of
motion which at least to high accuracy are time reversal invariant.
The ``essentially'' unique solution thus carries no time arrow
information.  From the point of view of this solution backward and
forward in time is quite on the same footing.  So from this point of
view it is equally miraculous that the brown mixed coffee and white cream
should ever have been separated as if it suddenly separated.
That is to say it would be a complete miracle if the ``essentially''
unique solution should lead to any irreversible processes as judged
on the macro level.  In fact we just have to invert the time axis and
remember that with inverted time axis an irreversible process looks as
a complete miracle caused by a completely unlikely micro
configuration.  

In this way we bring down our argument to just saying that the
``essentially'' unique solution is exceedingly unlikely to belong to the so
seldom class that it is -- time reversely looking -- a complete
miracle.  Compact space time gives so may equations of motion that the
solution becomes ``essentially'' unique.  That is, however,
incompatible with that the solution realized in nature with its
irreversible processes, is of a so miraculously seldom type, that 
we need the solution to have been chosen with such miracles (=the
irreversible processes) on purpose. 

One cannot get irreversible processes by accident, rather only by tuning
initial conditions just for that purpose.  Therefore the compact
space time is not allowed to fix the solution so that one (=God) cannot
fix it further so as to get the ``miraculous'' irreversible
processes.

\subsection{Configurations for which equations of motion progress a
chain}

In order to be able to apply our assumption of ``no miracles in either
way'' the easiest is to get hold of a configuration really a
co-dimension one series of lattice points the field values at which
one can deduce the field values at another similar configuration of
lattice points.  From the locality of the field equations which only
involve neighboring points it is rather obvious that to have
two hyper-lines or, curves or pieces of hyper-lines or curves able to
predict field values for each other they should together form a closed
loop, although they could possibly close at infinity.  Also it
is intuitively expected that the two hyper-curve pieces should be as
straight as possible.  Curving too much might get them correlated with
themselves. 
\footnote{
The method to find number of solutions of differential equations in
discretized space time is discussed in detail in [7] by the present
authors.}

It turned out as can be easily seen that drawing one of the
hyper-curves with full drawn line ----- and the other with
dashed line -\,-\,- the figures (see Fig. 1) of such 
series of lattice points related by equation of motion
can take the form:


\begin{figure}[h]
 \includegraphics[width=82mm]{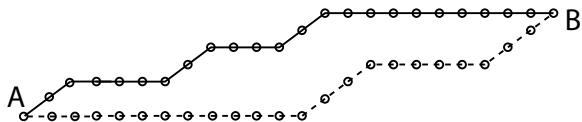}
 \caption{Hyper-curve on a lattice representing equation of motion}
\end{figure}

We consider here the two outer points A and B belonging to both
pieces of hyper-curves.  
In fact one can easily argue that if we consider the field value
sets on the dashed curve, by means of the equation of motion, 
then we can successively get determined
the field values inside the figure encircled by the two hyper-curve
pieces -\,-\,- and ----- and finally also get determined the field
values along the hyper-curve drawn in uninterrupted manner.  Presumably it is
the easiest to illustrate that one can in principle calculate the
mentioned field values by an example -- a parallelogram (Fig. 2)--.


\begin{figure}[h]
 \includegraphics[width=82mm]{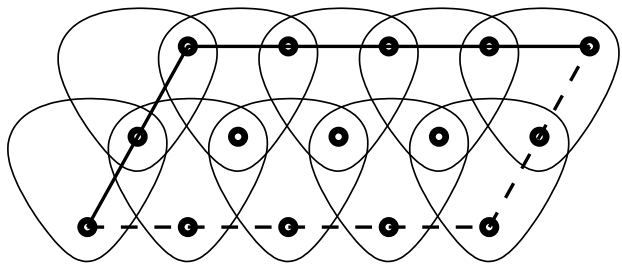}
 \caption{Parallelogram related with field values. 
 Inside triangles field components are calculable.}
\end{figure}

Suppose that we know the field components on the sites along the
interrupted curve (it is really here two lines with the angle
$120^\circ$ formed between them).  Let us also notice the rules for
calculability: Once we know two field component sets (i.e. the fields
on two sites) inside a set of three encircled sites by triangles 
$\bigtriangleup$ and $\bigtriangledown$, 
then we can compute the field components at the third of
those three sites uniquely.

Using this rule alone we can see, starting from the right most upgoing
part of the interrupted line series of sites, that we can uniquely 
calculate the field components at the line just one lattice constant further to the left 
parallel with this line.  Repeating the same
argument a step further to the left after that and continuing stepwise
one soon sees that we finally get in principle all the field
components on all the sites in the encircled parallelogram determined uniquely
by the two hyper-curve pieces.  Especially the sites on
the fully drawn curve piece get their field values determined. 

It is also easy to see that, if one oppositely started by knowing the
fields along the fully drawn line then one could determine all the
fields in the parallelogram.  This way one would, however, start by
determining the fields along the horizontal line of lattice points just
one lattice constant step below the uppermost side of the
parallelogram.

We may conceive of these two hyper curves with interrupted and full
drawn curves respectively as two space-like curves of general
relativity in two dimensions representing two different moments of
time.  To make this way of thinking more obvious we could let the two
pieces of hypercurves be continued in both directions but now
following each other totally (see Fig. 3):


\begin{figure}[h]
 \includegraphics[width=82mm]{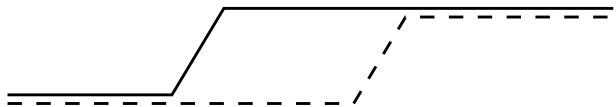}
 \caption{Example of the case in which two hyper-curves are interrupted.}
\end{figure}

Such a total following means that we have added chains of lattice
points to both hypercurves, the same lattice sites.  Since clearly a
point belonging to both is calculable from knowing either of the
hypercurves they do not play any role in the question of whether you
can or cannot calculate the fields on one hypercurve from that on the
other one.

But looking at it as two moments of time we can use the argumentation
above that successive entropies had to be equal -- i.e. entropy must be 
constant -- in the one-dimensional case, under the assumption of 
``reversability''.

If we by definition say that we shall use for the estimate of the
entropy at one of these moments of time only the entropies along the
hypercurve and the correlation entropy reductions between
neighboring sites along the hypercurve, we get expressions like 
\begin{eqnarray}
 \sum S_A + M_{A,~ A+1}.
\end{eqnarray}
Where $A$ is along hypercurve.
Here the $M_{AB}$ are expressions for the reduction in entropy of the
combined entropy of sites $A$ and $B$ due to their correlation.

In an example as Fig. 4 
we get for the entropy corresponding to the interrupted piece of curve
between $A$ and $H$:
\begin{figure}[h]
 \includegraphics[width=82mm]{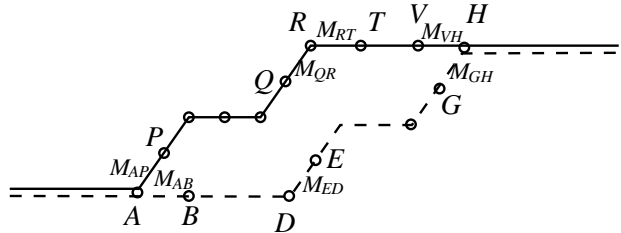}
 \caption{Example of entropy connected with field along the hyper-curve.}
\end{figure}

\begin{eqnarray}
S_{---} &=& S_{A} + M_{AB} + S_{B} + \ldots 
 + S_{D} +M_{DE}  
\nonumber\\&&
+S_{E}
+ \ldots 
 + S_{G} +M_{GH} +S_{H} 
\end{eqnarray}
This quantity of entropy is really the logarithm of the number of
different states in which the fields on the sites from $A$ to $H$ along
the interrupted line. 
This should be, as prescribed by the macro description, 
equal to the corresponding quantity for the fully drawn hypercurve 
by our ``reversability in either'' assumption.
The expression for the logarithm of the number of micro
states corresponding to the prescribed one is just given by
\begin{eqnarray}
 S \underline{~~~~~~} &=& S_{A} +M_{AP} + S_{P} + \ldots +S_{Q} +M_{QR} 
+ S_{R} \nonumber\\&&+
M_{RT} + S_{T} + \ldots 
+ S_{V}+ M_{VH} + S_{H}.
\end{eqnarray}

A trivial mathematical trick to simplify the notation is to absorb
the S-terms into the neighboring $M_{AB}$-terms by defining a quantity $K_{AB}$ for every
pair of neighboring sites on one of the two hypercurves considered
\begin{eqnarray}
 K_{AB}&\hat =&
  \frac{1}{2} S_{A} + \frac{1}{2} S_{B} + M_{AB} .
\end{eqnarray}
In this simplifying notation the equation 
$ S_{- - -} = S \underline{~~~~~~}$ or more precisely,
\begin{eqnarray}
&& S_{A} +M_{AB} + S_{B} + \ldots +S_{G} + M_{GH} + S_{H} 
 \nonumber\\&&
=
S_{A} +M_{AP} + S_{P} + \ldots +S_{V} + M_{VH} + S_{H}
\end{eqnarray}
becomes 
\begin{eqnarray}
 K_{AB} + \ldots + K_{GH} = K_{AP} + \ldots + K_{VH}.
\end{eqnarray}
We can conceive of this latter equation as describing a flow of
something -- in fact entropy -- crossing the links, $AB$ say, in amount
$K_{AB}$.  Then the equation we derived tells this something is
conserved.

Now, however, we have to discuss for which orientations of the used
parallelogram we shall consider the truncated and full hyper-curves
as tow different moments of times and thus should use our
argumentation that the development should not be reversable either way.

\subsection{Shall we assume no miraculous findings by moving in
space?}

Since we are anyway playing a mathematical game rather than working
viable physics when we play the game of throwing away the second law,
we may as well play several versions.

In the real world we are not surprised by high entropy density in some
places than in others.  It is quite non-miraculous that it is hotter at
some distance away from some other place, or that there could be 
ice and water may be of the same temperature but of different entropy
densities.  In real nature with the second law of thermodynamics a
temperature difference would not be obtainable in the long run and it
would be smoothened out.  
This means that a temperature difference
only in idealized models can be uphold from dismissing.  
Without second law and no miracles either we have only the one way out 
that there is constant temperature throughout the universe.  
The temperature difference growing would be miraculous and thus
should be excluded.  In time reversed way the smoothing out of
temperature would not be acceptable.  Except for possibly
several different phases realized in different regions we would thus
expect already the thermodynamical equilibrium that does not allow
temperature variations or chemical potential variations in space.  

Provided that there are some conserved quantities it should, however, still
be possible to have co-existing phases at the multiple point and thus 
it should \underline{not} be necessary 
to have the same entropy density all through space .  
But for such a situation to be upheld some conserved
charges $\partial_\mu j^\mu = 0$ would be needed.

At least with the possibility of having co-existing phases with all
chemical potentials and the temperatures in balance at some separating
border curve it seems that nothing should be miraculous locally at
all.  In as far as such a model should be realizable inside our very
general scheme, one would expect that we should find that a scheme
were realizable in our general scheme.

\section{Conclusion and Outlook}

We have studied the possibility of having a compact space time and
have come to the conclusion when using some very mild extra
assumptions that a compact space is not compatible with the
phenomenological fact that we have very often significant entropy
increase, irreversible processes.

We have used classical approximation -- and philosophy we could say --
all through, but honestly speaking if we have to rely on quantum
mechanics to get such numerically very big troubles as we estimate to
go away, it sounds very suspicious to us: Since our main point is to
count solutions, a passage through of the compact space time model by
use of quantum mechanics would easily come under the suspicion of
going through because one has forgotten that if you really get a very
tiny overlap for quantum probability it can mean that there is really
no overlap with the assumed wave function for the universe.

If one as it seems Hartle and Hawking do only use the wave function
for the universe as it comes out of the functional integral without
caring for if also the development into the future after the
measurement has been done using such a wave function, then there is
nothing that imposes a compact space time future.

Such a compact space time only to oneside in time is not in any
trouble of the type caused by totally compact space time as we
described it.  What would be impossible is only if one wants also to
have a no-boundary condition in the future. 

It is really the problem that one gets equally many equations of
motion as there are variables.  This means that it is the
problem to have irreversible processes in all models where there is
not some places or times where due to infinities or singularities or
for other reasons the number of equations compared to the number of
variables gets reduced.  But since we see the number of variables and
the number of equations just matching very nicely where we have the
phenomenological check, namely locally, it would be the most regular
and simplest behavior if it continued like that.  

A little possible addition to conclusion
may be interesting to mention in the following:
Since our argumentation is so strongly based on counting solutions
there is also the possibility of escaping our conclusion of no
irreversible processes even without disturbing the equality of the
number of equations of motion and variables to be determined from
them by the following loop hole: 

If there were in some way made an enormous number of ``attempts'' to
make a universe development -- a ``multiverse'' theory -- it could be
o.k. even if one could just find one solution for one of these
``attempts''.  In quantum mechanics in say the Feynman path way
formulations there \underline{is} in a formal way made such an enormous amount of
``attempts''.  In a quantum mechanical theory one could thus possibly
end up with a non zero but exceedingly small number of solutions per
attempt and still get formally a sensibly looking result.  From a
classical way of thinking an equation system with in first
approximation no solutions describing the world sounds ridiculous,  but in
a quantum formulation in which one always normalize distributions
before one confronts them with experiment such a naively seen nonsense
theory with an exceedingly unlikely universe may not be a true
problem.  

\begin{acknowledgments}
We acknowledge the Niels Bohr Institute (Copenhagen) and 
Yukawa Institute for Theoretical Physics for their hospitality
extended to one of them each.
The work is supported by Grant-in-Aids for Scientific Research 
on Priority Areas,  Number of Area 763 ``Dynamics of Strings and Fields", 
from the Ministry of Education of Culture, Sports, Science and Technology, Japan.
\end{acknowledgments}

\end{document}